\documentclass[prb, floatfix, twocolumn, superscriptaddress]{revtex4-2}
\def\be{\begin{equation}}
\def\ee{\end{equation}}
\def\bea{\begin{eqnarray}}
\def\eea{\end{eqnarray}}
\def\bi{\begin{itemize}}
\def\ei{\end{itemize}}
\usepackage{braket}
\usepackage{xcolor}
\usepackage{graphicx}
\usepackage{amsmath,amstext,amssymb,bm,color,times}
\usepackage[english]{babel}
\usepackage[T1]{fontenc}
\usepackage[utf8]{inputenc}
\usepackage[colorlinks=true,citecolor=blue,linkcolor=magenta]{hyperref}
\usepackage{float}

\begin{document}

%%%%%%%%%%%%%%%%%%%%%%%%%%%%%%%%%%%%%%%%%%%%%%%%%%%%%%%%%%%%%%%%%%%%%%%%%%

\title{Post-Critical Meson Dynamics of Kibble–Zurek Excitations in a 5,564-Qubit Quantum Annealer}

%%%%%%%%%%%%%%%%%%%%%%%%%%%%%%%%%%%%%%%%%%%%%%%%%%%%%%%%%%%%%%%%%%%%%%%%%%

\author{Francis A. Bayocboc Jr.}
\affiliation{Jagiellonian University, Institute of Theoretical Physics, ul. {\L}ojasiewicza 11, 30-348 Krak\'ow, Poland}
\author{Jacek Dziarmaga}
\affiliation{Jagiellonian University, Institute of Theoretical Physics, ul. {\L}ojasiewicza 11, 30-348 Krak\'ow, Poland}
\affiliation{Jagiellonian University,
             Mark Kac Center for Complex Systems Research,
             ul. \L{}ojasiewicza 11, 30-348 Krak\'ow, Poland}
\author{Marek M. Rams}
\affiliation{Jagiellonian University, Institute of Theoretical Physics, ul. {\L}ojasiewicza 11, 30-348 Krak\'ow, Poland}
\affiliation{Jagiellonian University,
             Mark Kac Center for Complex Systems Research,
             ul. \L{}ojasiewicza 11, 30-348 Krak\'ow, Poland}
\author{Jaka Vodeb}
\email{jaka.vodeb@ijs.si}
\affiliation{Department of Complex Matter, Jožef Stefan Institute, Jamova 39, 1000 Ljubljana, Slovenia}
\affiliation{CENN Nanocenter, Jamova 39, 1000 Ljubljana, Slovenia}
\affiliation{Jülich Supercomputing Centre, Institute for Advanced Simulation, Forschungszentrum Jülich, 52425 Jülich, Germany}

\date{\today}
%%%%%%%%%%%%%%%%%%%%%%%%%%%%%%%%%%%%%%%%%%%%%%%%%%%%%%%%%%%%%%%%%%%%%%%%%%

\begin{abstract}
Quantum phase transitions provide a controlled route for generating many-body excitations, but the dynamics after the critical point can be as important as the initial defect creation. Recent progress in quantum annealing has made it possible to access coherent nonequilibrium dynamics in programmable Ising systems with thousands of superconducting qubits. Here we use this capability to study a longitudinally biased quantum Ising chain, where Kibble--Zurek defect creation is followed by nonintegrable post-critical dynamics. The longitudinal bias confines kink--antikink excitations into mesonic bound states, so that the final spin configurations encode both the production of defects near the critical point and the subsequent evolution of the confined excitations. Using energy-scale rescaling and zero-noise extrapolation, we find that the defect density follows the expected biased Kibble--Zurek/Landau--Zener crossover and agrees with matrix-product-state simulations with uniform bias. In contrast, magnetization, spatial profiles, and minority-domain statistics reveal that mesonic evolution is interrupted by localization of the post-critical domain pattern. Matrix-product-state simulations with disorder reproduce this separation between robust defect creation and localized post-critical dynamics. Our results show that large-scale quantum annealers can probe the fate of critical excitations beyond defect counting.
\end{abstract}

\maketitle

%%%%%%%%%%%%%%%%%%%%%%%%%%%%%%%%%%%%%%%%%%%%%%%%%%%%%%%%%%%%%%%%%%%%%%%%%%%%
\section{Introduction}
\label{sec:introduction}
%%%%%%%%%%%%%%%%%%%%%%%%%%%%%%%%%%%%%%%%%%%%%%%%%%%%%%%%%%%%%%%%%%%%%%%%%%%%

Quantum phase transitions provide a central organizing principle for many-body physics. 
When a system is driven through a quantum critical point at a finite rate, the closing of the gap prevents fully adiabatic evolution and leads to the creation of defects according to the quantum Kibble--Zurek mechanism~\cite{QKZ1,QKZ2,QKZ3,d2005, K-a,*K-b,*K-c,Z-a,*Z-b,*Z-c, d2010-a,d2010-b,Z-d,ROSSINI20211}.
This mechanism has been explored across a broad range of platforms, including ultracold gases, strongly interacting Fermi superfluids, Rydberg-atom arrays, and programmable quantum annealers~\cite{ko2019kibble,keesling2019quantum,king2022coherent}. 
These studies have established defect formation and correlation-length scaling as robust probes of nonequilibrium critical dynamics.

The dynamics after the critical point can be substantially richer than the initial production of defects. 
In systems with broken discrete symmetry, the elementary excitations produced near the transition are domain walls. 
If a longitudinal field explicitly breaks the symmetry, isolated domain walls are no longer asymptotic quasiparticles: the field generates a string tension between a kink and an antikink and confines them into mesonic bound states~\cite{kormos2017real}. 
This confinement makes the post-critical dynamics nonintegrable and qualitatively changes relaxation. 
Theoretical work has shown that confined excitations can suppress correlation spreading and entanglement growth, generate long-lived coherent oscillations, stabilize prethermal regimes with approximately conserved meson number, and produce quasilocalized dynamics even in translation-invariant systems~\cite{kormos2017real,lerose2020quasilocalized,birnkammer2022prethermalization}. 
Related confinement physics has also been observed directly in quantum simulators, including trapped-ion spin chains and digital quantum circuits, where domain-wall bound states and confinement-induced changes in information and entanglement propagation were measured~\cite{tan2021domain,vovrosh2021confinement}.

These developments motivate going beyond defect counting. 
The kink density primarily probes excitation creation near the critical point, whereas magnetization profiles, domain-size statistics, and spatial correlations can retain information about the subsequent evolution of the confined excitations. 
Recent programmable quantum simulators have begun to access such post-critical dynamics, including coarsening and collective domain motion after quantum critical crossings~\cite{keesling2019quantum}. 
At the same time, quantum annealers have undergone a qualitative shift: coherent evolution through a quantum phase transition has been demonstrated in programmable Ising chains with thousands of superconducting flux qubits, and quantum critical dynamics has been extended to large spin-glass instances~\cite{king2022coherent,king2023quantum}. 
This opens a route to studying nonequilibrium dynamics in regimes where the relevant system sizes exceed those accessible to exact classical simulation, while still retaining a controlled connection to analytically and tensor-network tractable limits.

Here we use this capability to study a longitudinally biased one-dimensional quantum Ising chain implemented on a 5,564-qubit D-Wave quantum annealer. 
The longitudinal bias has two roles. 
First, it rounds the critical point and modifies Kibble--Zurek defect creation through the competition between the freeze-out length and the bias-induced magnetic length. 
Second, after the system enters the ordered phase, it confines the generated kink--antikink excitations into mesonic bound states whose subsequent evolution determines the final domain structure. 
We use energy-scale rescaling and zero-noise extrapolation to extract the quantum-simulation observables, and compare them with matrix-product-state simulations of both the uniform and disordered biased Ising chains.

Our results separate two stages of the dynamics. 
The final kink density follows the expected biased Kibble--Zurek/Landau--Zener crossover and agrees well with uniform-bias MPS simulations, showing that defect creation is governed primarily by near-critical physics. 
In contrast, the longitudinal magnetization, spatial profiles, and minority-domain statistics reveal post-critical dynamics beyond defect production. 
Uniform-bias MPS simulations show signatures of mesonic bound-state evolution, while the quantum-annealer data are consistent with an earlier localization of the post-critical domain pattern. 
This behavior is reproduced qualitatively by MPS simulations with static disorder in the longitudinal fields and nearest-neighbor couplings. 
The final spin configurations therefore encode both the universal creation of defects at a biased quantum critical point and the nonuniversal localization of the confined mesonic excitations formed after the crossing.

%%%%%%%%%%%%%%%%%%%%%%%%%%%%%%%%%%%%%%%%%%%%%%%%%%%%%%%%%%%%%%%%%%%%%%%%%%%%%%

%%%%%%%%%%%%%%%%%%%%%%%%%%%%%%%%%%%%%%%%%%%%%%%%%%%%%%%%%%%%%%%%%%%%%%%%%%%%
\section{Quantum simulation of a biased quantum Ising chain}
\label{sec:annealer_realization}
%%%%%%%%%%%%%%%%%%%%%%%%%%%%%%%%%%%%%%%%%%%%%%%%%%%%%%%%%%%%%%%%%%%%%%%%%%%%

\begin{figure}[t]
\centering
\includegraphics[width=\linewidth]{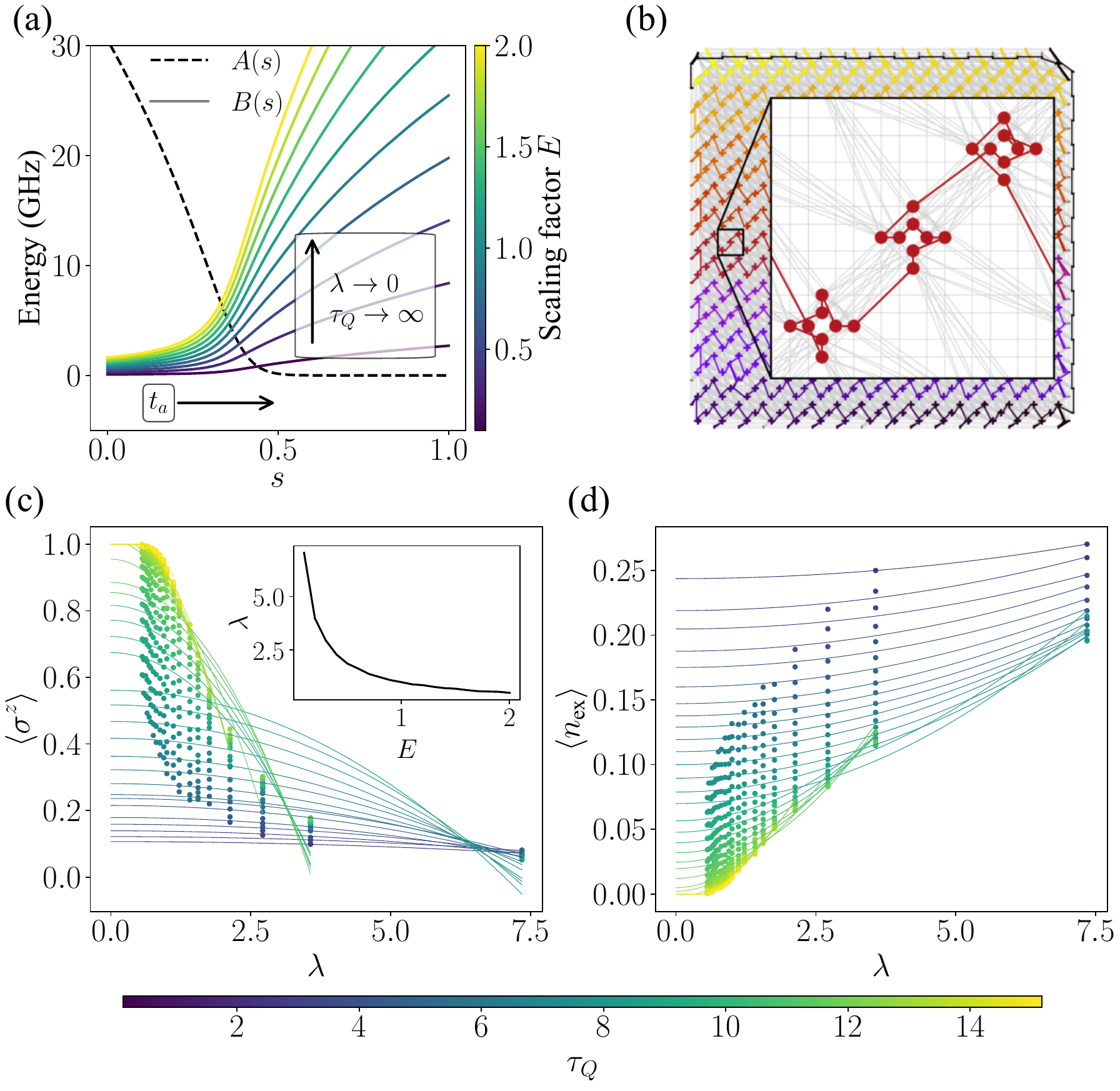}
\caption{
\textbf{Quantum simulation protocol and zero-noise extrapolation.}
(a) D-Wave fast-annealing schedule, showing the transverse-field energy scale $A(s)$ and Ising energy scale $B(s)$ as functions of the annealing parameter $s$. The colored curves show the effective rescaling $B(s)\rightarrow E B(s)$ used for zero-noise extrapolation. Increasing $E$ shifts the crossing toward smaller effective noise $\lambda$, while the physical annealing time $t_a$ sets the quench time $\tau_Q$ near the critical point.
(b) Embedding of the one-dimensional Ising chain on the D-Wave quantum processing unit. The red path denotes the active 5,564-qubit chain used in the experiment.
(c),(d) Representative zero-noise extrapolations at longitudinal bias $h=0.1$ for the final magnetization $\langle \sigma^z\rangle$ and defect density $\langle n_{\rm ex}\rangle$, respectively. Points show quantum-annealer data obtained at different energy scales $E$, expressed through the dimensionless noise parameter $\lambda$; solid curves are fits including constant and quadratic terms in $\lambda$. The intercept at $\lambda=0$ gives the zero-noise estimate. The color scale denotes the corresponding quench time $\tau_Q$. The inset in (c) shows the mapping between the hardware scaling factor $E$ and the extrapolation parameter $\lambda$.
}
\label{fig:1}
\end{figure}

The Hamiltonian we implemented on the D-Wave Advantage ($\mathrm{Advantage\_system5\_4}$ and $\mathrm{Advantage\_system6\_4}$) quantum annealer is
\begin{equation}
H_{DW} =
-\Gamma(s) \sum_{i=1}^N\sigma_i^x
-{\cal J}(s)
\left(
h\sum_{i=1}^N\sigma_i^z +
J\sum_{\langle i<j\rangle}^N\sigma_i^z\sigma_j^z
\right),
\label{eq:HDW}
\end{equation}
where $\sigma_i^{x}$ and $\sigma_i^{z}$ denote Pauli matrices. The functions $\Gamma(s)=A(s)/2$ and ${\cal J}(s)=B(s)/2$ are time-dependent energy prefactors that evolve during the annealing schedule, as shown in Fig.~\ref{fig:1}(a). Here, $J=1$ denotes the nearest-neighbor Ising coupling along the embedded chain, while $h\geq 0$ is a uniform longitudinal bias field. For $h=0$ and unit energy scale, the annealing ramp traverses the quantum critical point of the one-dimensional transverse-field Ising chain at $s_c$, where $A(s_c)/B(s_c)=1$. The embedding of the $N=5564$ qubit chain onto the D-Wave quantum processing unit is shown in Fig.~\ref{fig:1}(b).

%%%%%%%%%%%%%%%%%%%%%%%%%%%%%%%%%%%%%%%%%%%%%%%%%%%%%%%%%%%%%%%%%%%%%%%%%%%%%%
%%%%%%%%%%%%%%%%%%%%%%%%%%%%%%%%%%%%%%%%%%%%%%%%%%%%%%%%%%%%%%%%%%%%%%%%%%%%%%

The physical implementation is not free from dissipation and decoherence. However, as shown in Ref.~\onlinecite{dwave_mitigation} using a Bloch-Redfield description, these effects can be substantially mitigated for annealing across a quantum critical point. This mitigation procedure does not correct imperfections in the implemented Hamiltonian \eqref{eq:HDW}, such as static deviations in local fields or couplings, which will become important for the post-critical dynamics discussed below.

For the purpose of error mitigation, the longitudinal field and Ising coupling in $H_{DW}$ can be rescaled by multiplying $h$ and $J$ by a common prefactor $E$, constrained by the hardware to the range $0.1\leq E\leq 2$. The lower bound corresponds to a strongly reduced interaction scale, for which noise dominates the dynamics and the system behavior becomes nearly indistinguishable from the case $E=0$. The upper bound is imposed by the hardware limits of the quantum processing unit. Varying $E$ effectively rescales the annealing schedule according to $B(s)\rightarrow E B(s)$.

The critical point of the rescaled Hamiltonian is therefore shifted to the value $s_c=s_c(E)$ determined by
\begin{equation}
A(s_c)=E B(s_c).
\end{equation}
Following Ref.~\onlinecite{dwave_mitigation}, we define the dimensionless noise parameter
\begin{equation}
\lambda = t_Q(E)/t_Q(1),
\end{equation}
where
\begin{equation}
t_Q(E)=
\frac{g_c}{\pi E B(s_c)}
\left[
\frac{B'(s_c)}{B(s_c)}-\frac{A'(s_c)}{A(s_c)}
\right]
\label{eq:tQ}
\end{equation}
is the characteristic time scale of the annealing ramp near the critical point. Here all schedule functions and derivatives are evaluated at the rescaled critical point $s_c(E)$. In the limit $\lambda\rightarrow 0$, the system approaches the zero-noise regime. The dependence of $\lambda$ on the scaling factor $E$ is shown in the inset of Fig.~\ref{fig:1}(c), while representative zero-noise extrapolations of the quantum-simulation observables are shown in Figs.~\ref{fig:1}(c,d).

The D-Wave annealing schedule is nonlinear in the parameter $s$. Since the critical dynamics is controlled by the vicinity of $s_c$, we approximate the schedule locally by a linear crossing of the critical point and define the corresponding dimensionless quench time as
\begin{equation}
\tau_Q=t_a/t_Q(E),
\label{eq:tauQ}
\end{equation}
where $t_a$ is the duration of a linear quench $s(t)=t/t_a$ from $0$ to $1$. The two independent control parameters used below are therefore the longitudinal bias $h$, which controls confinement and the bias-induced critical length scale, and the quench time $\tau_Q$, which controls the Kibble--Zurek freeze-out length. For the full description of the parameter regimes used in our quantum simulation, see App.~\ref{app:parameter_sweep}. 

In the next section we first use the final kink density to establish the near-critical production of excitations in the biased Ising chain. We then turn to magnetization, domain-size statistics, and spatially resolved profiles, which probe the post-critical evolution and localization of the confined excitations.

%%%%%%%%%%%%%%%%%%%%%%%%%%%%%%%%%%%%%%%%%%%%%%%%%%%%%%%%%%%%%%%%%%%%%%%%%%%%
\section{Kibble--Zurek defect creation in a longitudinal bias}
\label{sec:kz_bias}
%%%%%%%%%%%%%%%%%%%%%%%%%%%%%%%%%%%%%%%%%%%%%%%%%%%%%%%%%%%%%%%%%%%%%%%%%%%%

%%%%%%%%%%%%%%%%%%%%%%%%%%%%%%%%%%%%%%%%%%%%%%%%%%%%%%%%%%%%%%%%%%%%%%%%%%%%%%
\begin{figure}[t]
\centering
\includegraphics[width=\linewidth]{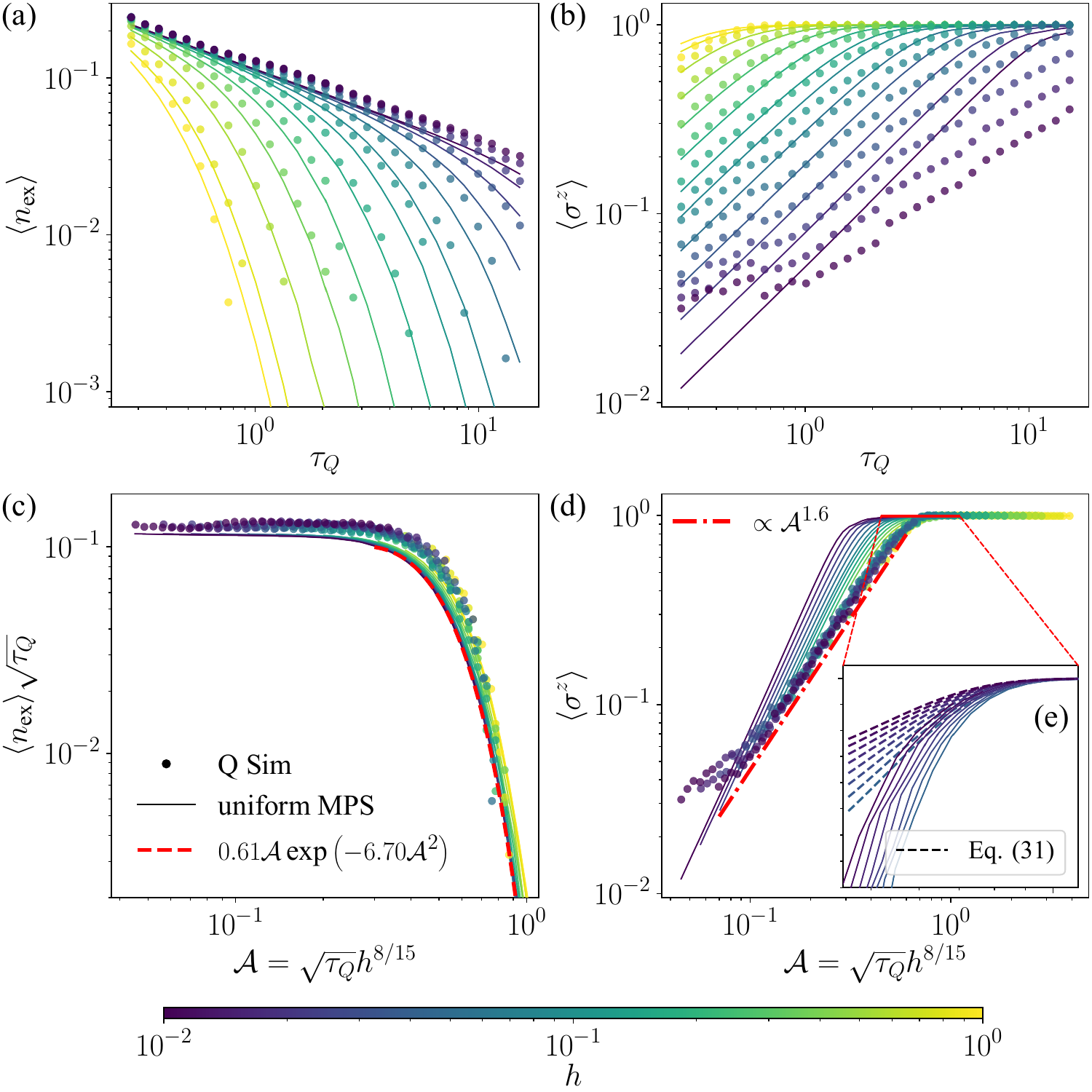}
\caption{
\textbf{Biased Kibble--Zurek scaling and post-critical magnetization.}
(a),(b) Zero-noise--extrapolated quantum simulation data (dots) for the final defect density $\langle n_{\rm ex}\rangle$ and magnetization $\langle\sigma^z\rangle$, compared with uniform-bias MPS simulations (solid lines), as functions of the quench time $\tau_Q$ for different longitudinal biases $h$.
(c) Defect-density data rescaled by the Kibble--Zurek length $\hat\xi\propto\sqrt{\tau_Q}$ and plotted against the biased scaling variable ${\cal A}=\hat\xi/\xi_{\|}\propto\sqrt{\tau_Q}h^{8/15}$. The collapse shows the crossover from Kibble--Zurek defect creation at small ${\cal A}$ to Landau--Zener suppression at large ${\cal A}$; the dashed line is a fit of the form $\alpha{\cal A}\exp(-a{\cal A}^2)$.
(d) Final magnetization plotted against the same scaling variable. The quantum-simulation data collapse approximately as a function of ${\cal A}$ and are well described at small and intermediate ${\cal A}$ by a power law $\propto{\cal A}^{1.6\pm0.01}$, consistent with an early freezing of the bias-induced domain-length imbalance. The uniform-bias MPS simulations show an additional spread at large magnetization because, in the uniform system, post-critical meson dynamics continues after the critical crossing. The inset compares this spread with Eq.~\eqref{eq:Mf_clean_adiab}, which uses the adiabatic relation $M_f\simeq1-n_{\rm ex}$ together with the Landau--Zener estimate for the density of lowest mesonic excitations; the remaining deviation at smaller ${\cal A}$ occurs because higher mesonic states and nonadiabatic Kibble--Zurek dynamics invalidate the lowest-meson adiabatic approximation.
}
\label{fig:ZNE_M_kink}
\end{figure}
%%%%%%%%%%%%%%%%%%%%%%%%%%%%%%%%%%%%%%%%%%%%%%%%%%%%%%%%%%%%%%%%%%%%%%%%%%%%%%

The annealing dynamics across a quantum critical point is described by the quantum version~\cite{QKZ1,QKZ2,QKZ3,d2005} of the Kibble--Zurek mechanism~\cite{K-a,*K-b,*K-c,Z-a,*Z-b,*Z-c}; see also the reviews in Refs.~\cite{d2010-a,d2010-b,Z-d,ROSSINI20211}. A system initially prepared in its ground state is driven across the critical point at a finite rate. Far from criticality the evolution is adiabatic, but near the critical point the closing of the many-body gap prevents the state from following the instantaneous ground state. The resulting freeze-out length controls the density of defects produced during the crossing.

Close to the critical point, the distance from criticality can be linearized as
\begin{equation}
\epsilon(t)=(t-t_c)/\tau_Q,
\end{equation}
where $\tau_Q$ is the quench time. Adiabaticity breaks down at $t_c-\hat t$, where
\begin{equation}
\hat t \propto \tau_Q^{z\nu/(1+z\nu)},
\label{eq:hatt}
\end{equation}
and is restored after the system leaves the critical region at $t_c+\hat t$. The correlation length inherited from the ground state at the freeze-out time is
\begin{equation}
\hat\xi \propto \tau_Q^{\nu/(1+z\nu)}.
\label{eq:hatxi_general}
\end{equation}
For the one-dimensional quantum Ising transition, which belongs to the two-dimensional classical Ising universality class, the critical exponents are
\begin{equation}
z=1,\qquad \nu=1,\qquad \beta=\frac18,\qquad \gamma=\frac74,\qquad \delta=15.
\end{equation}
Therefore,
\begin{equation}
\hat\xi \propto \sqrt{\tau_Q}.
\label{eq:hatxi}
\end{equation}
The scale $\hat\xi$ sets the characteristic distance between kinks produced during the crossing, and hence the leading scaling of the final defect density. Other observables, however, can remain sensitive to post-critical evolution in the ordered phase~\cite{QKZteor-e,RadekNowak,dziarmaga_kinks_2022,dwave_mitigation}.

The longitudinal bias introduces a second length scale and modifies the critical crossing. At the critical point, a finite longitudinal field rounds the transition. The equilibrium magnetization induced by the field scales as
\begin{equation}
M_c \propto h^{1/\delta}=h^{1/15},
\end{equation}
while the correlation length and gap become finite,
\begin{equation}
\xi_{\|} \propto h^{-\nu/\beta\delta}=h^{-8/15},
\end{equation}
and
\begin{equation}
\Delta_{\|} \propto h^{z\nu/\beta\delta}=h^{8/15}.
\end{equation}
The relevant dimensionless parameter is the ratio between the Kibble--Zurek length $\hat\xi$ and the bias-induced magnetic length $\xi_{\|}$:
\begin{equation}
\hat\xi/\xi_{\|} \propto \tau_Q^{1/2} h^{8/15} \equiv {\mathcal A}.
\label{eq:adiab_cond}
\end{equation}
For ${\mathcal A}\ll1$, the system crosses the critical region in the Kibble--Zurek regime: the bias is weak on the scale of the freeze-out length. For ${\mathcal A}\gg1$, the bias-induced gap is large enough to suppress defect creation, and the dynamics crosses over toward an effectively adiabatic Landau--Zener regime. The corresponding excitation probability is~\cite{QKZteor-r}
\begin{equation}
p = e^{-a \tau_Q h^{16/15}} \equiv e^{-a {\mathcal A}^2},
\label{eq:p}
\end{equation}
where $a=6.89$ was estimated from MPS simulations~\cite{QKZteor-r}.

Figure~\ref{fig:ZNE_M_kink}(a,c) compares the zero-noise--extrapolated quantum simulation results, uniform-bias MPS simulations, and the biased Kibble--Zurek/Landau--Zener scaling prediction for the final kink density. The agreement is good over the range of biases and quench times studied here. After rescaling by the Kibble--Zurek length, the data collapse onto a single scaling function of ${\mathcal A}$. This shows that the final density of kinks is primarily controlled by defect creation near the critical point.

The longitudinal magnetization contains complementary information. Unlike the kink density, which counts the number of domain walls, the magnetization also depends on the relative lengths of domains aligned and anti-aligned with the bias. It is therefore sensitive to how the domain pattern evolves after the critical crossing. Figure~\ref{fig:ZNE_M_kink}(b,d) shows the final magnetization $M=\langle\sigma^z\rangle$. In the limit ${\mathcal A}\gg1$, defect creation is strongly suppressed and the magnetization saturates at $M\simeq1$. In the opposite limit ${\mathcal A}\ll1$, the dynamics is still governed by Kibble--Zurek freeze-out, but the longitudinal field leaves an imprint on the magnetization.

This imprint can be estimated from the equilibrium susceptibility at the freeze-out point. At $t_c-\hat t$, the field-induced magnetization is
\begin{equation}
\hat M \propto h \tau_Q^{\gamma/(1+z\nu)} = h\tau_Q^{7/8}.
\end{equation}
It is carried through the impulse regime until adiabaticity is restored at $t_c+\hat t$. Equivalently, its value $\hat M_c$ at the critical point obeys~\cite{QKZteor-r}
\begin{equation}
\hat M_c/M_c \propto
\hat M/M_c \propto
(\hat\xi/\xi_{\|})^{\gamma/\nu}
=
{\mathcal A}^{7/4}.
\end{equation}
In practice the magnetization does not remain strictly frozen during the impulse regime. Critical ordering continues to amplify it, so that after the crossing the magnetization is $\kappa\hat M$, with $\kappa>1$, see App.~\ref{app:sonic_M}.

It is useful to translate this statement into a domain picture. Shortly after the critical point, the system consists of a mosaic of ferromagnetic domains. Locally, the ordered moment in domains aligned and anti-aligned with the bias is approximately $\pm\hat M_s$, with a small field-induced shift $\hat M$. If the two types of domains occupied equal fractions of the system, their spontaneous contributions would cancel and the net magnetization would be set only by the field-induced shift. The fact that the magnetization is enhanced to $\kappa\hat M$ means that domains aligned with the bias must occupy a larger fraction of the chain than domains anti-aligned with it.

Let the imbalance between these length fractions be denoted by $x$. The excess magnetization beyond the uniform field-induced shift is then represented by
\begin{equation}
x=(\kappa-1)\hat M/\hat M_s.
\end{equation}
Using $\hat M\propto h\tau_Q^{7/8}$ and $\hat M_s\propto \tau_Q^{-\beta/(1+z\nu)}=\tau_Q^{-1/16}$ gives
\begin{equation}
x\propto h\tau_Q^{15/16}\propto {\mathcal A}^{15/8}.
\end{equation}
Thus, Kibble--Zurek freeze-out in a longitudinal bias not only creates domain walls, but also imprints a bias-dependent imbalance between the lengths of the two symmetry-broken domain orientations.

This observation provides a simple interpretation of the quantum simulation data. Suppose that, shortly after $t_c+\hat t$, the positions of the domain walls become effectively localized and remain fixed until the end of the ramp. The local ordered moment within each domain can still grow adiabatically toward its final value, but the relative domain lengths no longer change. In that case the final magnetization is controlled by the frozen length imbalance,
\begin{equation}
M_f \simeq x \propto {\mathcal A}^{15/8}.
\end{equation}
In Fig.~\ref{fig:ZNE_M_kink}(d), the zero-noise--extrapolated quantum simulation data collapse approximately as a function of ${\mathcal A}$ and are fitted by $M_f\propto{\mathcal A}^{1.6}$. The fitted exponent differs from the crude estimate $15/8=1.875$, but the collapse itself is the important point: it suggests that the final magnetization is largely determined by the domain imbalance generated near the critical point.

MPS simulations behave differently. Although their kink density follows the same biased Kibble--Zurek scaling, their magnetization does not collapse perfectly as a function of ${\mathcal A}$. Instead, the rescaled MPS data show a residual spread. This spread indicates that, in the uniform biased Ising chain, the domains continue to evolve after the critical region is crossed. In particular, the confined kink--antikink excitations generated near the transition undergo post-critical meson dynamics, whose details depend not only on ${\mathcal A}$ but also on the separate values of $h$ and $\tau_Q$.

The comparison therefore separates two pieces of physics. The kink density measures the near-critical production of defects and follows the biased Kibble--Zurek/Landau--Zener scaling. The magnetization probes the subsequent evolution of the domain pattern. Its collapse in the quantum simulation suggests early freezing of the post-critical domain structure, while the residual spread in the MPS data points to continued mesonic evolution in the ideal system. In the next section we show that static disorder in the longitudinal fields and nearest-neighbor couplings naturally produces such freezing by localizing the post-critical domain walls.

%%%%%%%%%%%%%%%%%%%%%%%%%%%%%%%%%%%%%%%%%%%%%%%%%%%%%%%%%%%%%%%%%%%%%%%%%%%%
\section{Localization of the post-critical domain pattern}
\label{sec:localization}
%%%%%%%%%%%%%%%%%%%%%%%%%%%%%%%%%%%%%%%%%%%%%%%%%%%%%%%%%%%%%%%%%%%%%%%%%%%%

The collapse of the final magnetization as a function of ${\mathcal A}$ suggests that the domain pattern in the quantum simulation becomes effectively frozen shortly after the critical region. A natural mechanism for such freezing is static disorder in the implemented longitudinal fields and nearest-neighbor couplings. Such disorder does not primarily change the universal defect creation mechanism near the critical point, but it can strongly affect the subsequent motion of domain walls in the ordered phase.

To test this interpretation, we compare the zero-noise--extrapolated quantum simulation data with MPS simulations of a disordered biased Ising chain. In these simulations, the uniform Hamiltonian \eqref{eq:HDW} is modified by allowing both the local bias and the nearest-neighbor coupling to fluctuate around their target values,
\begin{equation}
\begin{aligned}
H_{\rm dis} =
&-\Gamma(s) \sum_i \sigma_i^x \\
&-{\mathcal J}(s)
\left[
\sum_i \left(h+\delta h_i\right)\sigma_i^z
+
\sum_i \left(J+\delta J_i\right)\sigma_i^z\sigma_{i+1}^z
\right].
\end{aligned}
\label{eq:Hdis}
\end{equation}
The disorder variables are taken to be Gaussian and statistically independent, with zero mean and variances set by $\sigma_h$ and $\sigma_J$. In the comparison below we use $\sigma_h=\sigma_J=\sigma$, unless stated otherwise. The disorder is static during a ramp and therefore represents Hamiltonian imperfections rather than dynamical decoherence.

%%%%%%%%%%%%%%%%%%%%%%%%%%%%%%%%%%%%%%%%%%%%%%%%%%%%%%%%%%%%%%%%%%%%%%%%%%%%%%
\begin{figure}[t]
\centering
\includegraphics[width=\linewidth]{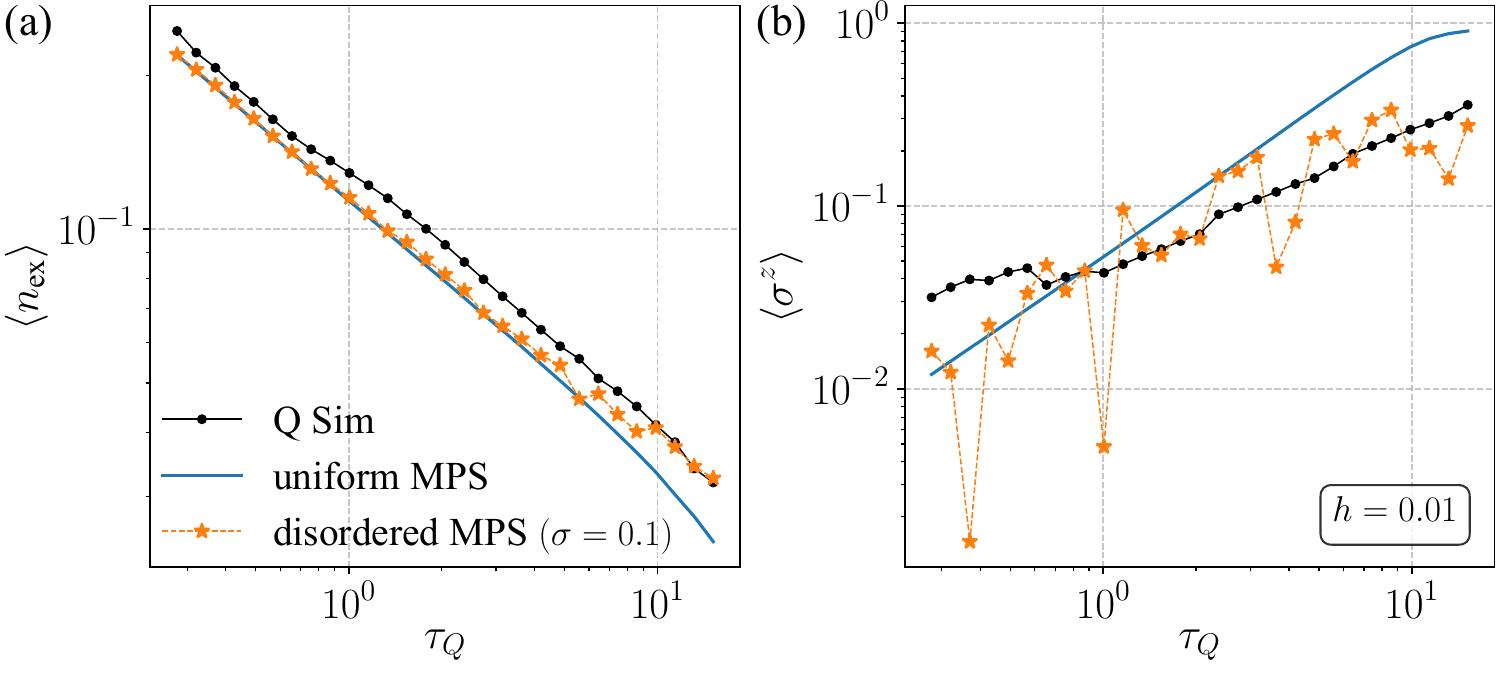}
\caption{
\textbf{Effect of static disorder on defect creation and magnetization.}
Zero-noise--extrapolated quantum simulation data (black dots) at $h=0.01$, compared with clean uniform-bias MPS simulations (solid blue line) and disordered MPS simulations (orange stars).
(a) Final defect density $\langle n_{\rm ex}\rangle$ as a function of quench time $\tau_Q$. The disordered MPS data remain close to both the uniform-bias MPS and quantum-simulation results, showing that the total number of defects is primarily set by near-critical Kibble--Zurek creation and is comparatively insensitive to weak static disorder. The remaining disparity is likely due to thermal effects on the device, not taken into account by our numerics.
(b) Final magnetization $\langle\sigma^z\rangle$ for the same parameters. In contrast to the defect density, the magnetization is strongly modified by disorder and shifts toward the quantum-simulation data. This separation between robust defect density and disorder-sensitive magnetization indicates that static disorder mainly affects the post-critical domain evolution rather than the initial creation of defects. The large fluctuations are due to a relatively small amount of disorder-instance averaging in MPS ($\sim10$) compared to the large number of samples gathered in quantum simulations ($\sim10^6$).
}
\label{fig:ZNE_M_kink_vs_noisy_MPS}
\end{figure}
%%%%%%%%%%%%%%%%%%%%%%%%%%%%%%%%%%%%%%%%%%%%%%%%%%%%%%%%%%%%%%%%%%%%%%%%%%%%%%

Figure~\ref{fig:ZNE_M_kink_vs_noisy_MPS} shows the comparison for $h=0.01$. The defect density is only weakly affected by the disorder over a broad range of annealing times. This is consistent with the interpretation that $n_{\rm ex}$ is mainly determined by defect creation near the critical point, where the relevant scale is the Kibble--Zurek length $\hat\xi$. By contrast, the final magnetization is much more sensitive to disorder. While the MPS result retains the post-critical evolution of the uniform biased Ising chain, the disordered MPS result is shifted toward the quantum simulation data. Thus, the same disorder that leaves the kink density approximately intact can substantially modify the magnetization by freezing or pinning the post-critical domain pattern.

This behavior is expected because the final magnetization depends not only on how many kink--antikink pairs are produced, but also on how far the domain walls move after their production and how the relative lengths of domains aligned and anti-aligned with the bias evolve during the remainder of the ramp. Static disorder introduces preferred positions for domain walls. Once the transverse field is sufficiently small, domain-wall motion becomes ineffective, and the final magnetization reflects a pinned spatial configuration rather than post-critical dynamics in the uniform bias case.

\begin{figure}[t]
\centering
\includegraphics[width=\linewidth]{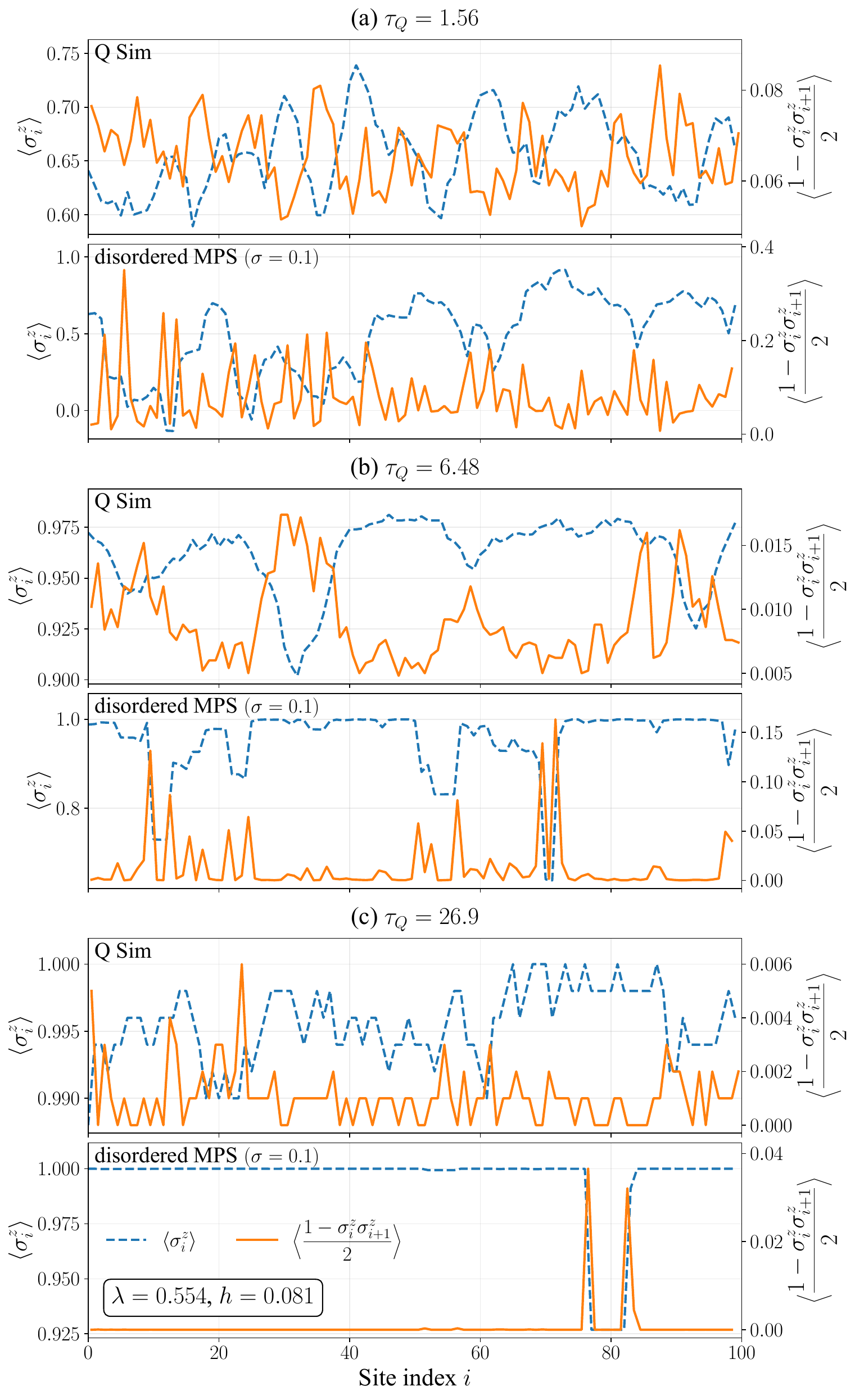}
\caption{
\textbf{Spatial signatures of post-critical domain-wall localization.}
Local magnetization $M_i=\langle\sigma_i^z\rangle$ (blue dashed lines) and local defect density
$n_i=\langle(1-\sigma_i^z\sigma_{i+1}^z)/2\rangle$ (orange solid lines) at $h=0.081$, comparing zero-noise--extrapolated quantum simulation data with disordered MPS simulations for increasing quench times:
(a) $\tau_Q=1.56$, (b) $\tau_Q=6.48$, and (c) $\tau_Q=26.9$.
In both the quantum simulation and disordered MPS data, the local defect density is concentrated at preferred positions rather than being spatially uniform. These peaks coincide with sharp changes in the magnetization profile, showing that the final kinks act as pinned domain walls separating locally ordered regions. The persistence of this structure with increasing $\tau_Q$ supports the interpretation that static disorder localizes the post-critical domain pattern after Kibble--Zurek defect creation.
}
\label{fig:ZNE_M_kink_profile_vs_noisy_MPS}
\end{figure}

The spatial profiles provide direct evidence for this interpretation. In Fig.~\ref{fig:ZNE_M_kink_profile_vs_noisy_MPS}, we compare local magnetization and local defect-density profiles from the quantum simulation and from disordered MPS simulations. The local defect density is defined on bonds as
\begin{equation}
n_i=\left\langle \frac{1-\sigma_i^z\sigma_{i+1}^z}{2}\right\rangle .
\end{equation}
A nonzero value of $n_i$ identifies the presence of a kink on the bond between sites $i$ and $i+1$.

%%%%%%%%%%%%%%%%%%%%%%%%%%%%%%%%%%%%%%%%%%%%%%%%%%%%%%%%%%%%%%%%%%%%%%%%%%%%%%
%%%%%%%%%%%%%%%%%%%%%%%%%%%%%%%%%%%%%%%%%%%%%%%%%%%%%%%%%%%%%%%%%%%%%%%%%%%%%%

The profiles show that the final kinks are not distributed uniformly. Instead, they preferentially appear at specific positions, where they are correlated with abrupt variations in the magnetization profile. This behavior is reproduced qualitatively by the disordered MPS simulations. The agreement is not expected to be one-to-one at the level of individual sites, since the exact microscopic disorder profile in the quantum annealer is not independently known. The relevant point is that static disorder produces the same qualitative phenomenology: domain walls become pinned at preferred positions, and the magnetization changes discontinuously across those pinned locations.

The disorder therefore separates two aspects of the dynamics. The creation of kink excitations remains governed mainly by the biased Kibble--Zurek/Landau--Zener mechanism described in Sec.~\ref{sec:kz_bias}. The subsequent post-critical evolution of the domain pattern, however, is sensitive to static Hamiltonian imperfections. In the uniform system, the confined kink--antikink excitations continue to evolve as mesonic bound states after the critical crossing. In the disordered system, this post-critical meson dynamics is interrupted by localization of the domain walls. The next section develops the mesonic picture without disorder and shows how its domain-size predictions are modified by localization.

%%%%%%%%%%%%%%%%%%%%%%%%%%%%%%%%%%%%%%%%%%%%%%%%%%%%%%%%%%%%%%%%%%%%%%%%%%%%
\section{Post-critical meson dynamics without disorder}
\label{sec:meson_dynamics}
%%%%%%%%%%%%%%%%%%%%%%%%%%%%%%%%%%%%%%%%%%%%%%%%%%%%%%%%%%%%%%%%%%%%%%%%%%%%

The localization discussed in Sec.~\ref{sec:localization} interrupts the post-critical dynamics of the uniform biased Ising chain. To identify what is being interrupted, we now describe the corresponding evolution in the case without disorder. Without disorder, the post-critical dynamics is not negligible, even deep in the adiabatic regime where the excitation density remains low. To zeroth order in small $h$, the excitations consist of pairs of fermionic quasiparticles in the ferromagnetic phase, i.e., kinks dressed by quantum fluctuations induced by the transverse field. To first order in $h$, the longitudinal bias generates a linear confining potential between a kink and an antikink, binding them into mesonic eigenstates~\cite{Lagnese_Wilczek}. The low-energy dynamics is dominated by the lowest mesonic states.

In principle, these states can decay into multiple mesons via string breaking. However, this process is negligibly slow on the time scales considered here, and the number of quasiparticles is approximately conserved. Consequently, the final kink density remains approximately equal to the density of Kibble--Zurek excitations generated shortly after crossing the critical point. This explains why the defect density is well described by the biased Kibble--Zurek/Landau--Zener scaling of Sec.~\ref{sec:kz_bias}, even though other observables remain sensitive to the post-critical evolution.

The excited mesons continue to evolve as $\Gamma$ is ramped down to zero, see App.~\ref{app:mesons} for a detailed derivation. Initially, their dynamics is adiabatic due to the gaps separating mesonic states, which scale as
\begin{equation}
\Delta_{\rm meson} \propto \Gamma^{1/3}({\mathcal J}h)^{2/3}.
\label{eq:Deltam}
\end{equation}
As $\Gamma$ decreases, the meson size shrinks adiabatically according to
\begin{equation}
\xi_{\rm meson} \propto \Gamma^{1/3} \left({\mathcal J}h\right)^{-1/3}.
\label{eq:xim}
\end{equation}
Given that the transition rate scales as $\propto t_a^{-1}$, adiabaticity breaks down when $\Delta_{\rm meson}\sim t_a^{-1}$, at which point the meson size freezes at
\begin{equation}
\tilde \xi \propto \left({\mathcal J}_f h t_a\right)^{-1}.
\label{eq:tildexi_main}
\end{equation}
This frozen meson size sets the typical size of the spin-down, or false-vacuum, domains at the end of the ramp.

For longer annealing times, when the predicted $\tilde\xi$ approaches unity or becomes smaller, the ramp remains adiabatic until its end. In this regime, instead of freezing out at a finite continuum size, the mesonic states evolve adiabatically into excitations characteristic of a weak transverse field, $\Gamma\ll{\mathcal J}h$. These excitations correspond to spin-up domains of various lengths $L$ separated by energy gaps $\propto{\mathcal J}h$. The $l$-th mesonic state evolves into a domain of size $L=l$, and the probability of observing a domain of size $L$ is determined by the excitation probability of the $l$-th state near the critical point.

In the adiabatic regime of sufficiently large ${\mathcal A}$, only the lowest mesonic states, $l=1$, are likely to be excited near the critical point, with probability
\begin{equation}
p_1=p=e^{-a{\mathcal A}^2}.
\end{equation}
They subsequently evolve adiabatically into single reversed spins. Consequently, the final magnetization is related to the final kink density as
\begin{equation}
M_f \approx 1 - n_{\rm ex}.
\label{eq:Mf_adiab}
\end{equation}

More generally, the final density of kinks is expected to satisfy
\begin{equation}
\hat\xi\, n_{\rm ex} =
F\left(\hat\xi,\hat\xi/\xi_{\|}\right),
\label{eq:F_general}
\end{equation}
where $F$ is a nonuniversal scaling function. For sufficiently small ${\cal A}\equiv\hat\xi/\xi_{\|}$, one recovers the Kibble--Zurek regime with $F=\mathrm{const}$; see Fig.~\ref{fig:ZNE_M_kink}(c). In the complementary adiabatic regime,
\begin{equation}
n_{\rm ex} =
\alpha \xi_{\|}^{-1} e^{-a{\cal A}^2}.
\label{eq:nex_adiab}
\end{equation}
Here, the exponent is the Landau--Zener excitation probability in Eq.~\eqref{eq:p}. The magnetic length $\xi_{\|}=h^{-8/15}$ characterizes the width of the lowest excitation near the critical point, while $\xi_{\|}^{-1}$ sets the corresponding linear density of orthogonal excitation states. Multiplying Eq.~\eqref{eq:nex_adiab} by the Kibble--Zurek length gives
\begin{equation}
\hat\xi\, n_{\rm ex}
=
\alpha{\cal A}e^{-a{\cal A}^2},
\label{eq:F_adiab}
\end{equation}
where the proportionality constant in $\hat\xi\propto\sqrt{\tau_Q}$ has been absorbed into $\alpha$. This form provides a natural fitting function for Fig.~\ref{fig:ZNE_M_kink}(c), with two fitting parameters, $\alpha$ and $a$.

Combining Eqs.~\eqref{eq:Mf_adiab} and \eqref{eq:nex_adiab}, we obtain the uniform-bias prediction for the magnetization at the end of the ramp:
\begin{equation}
M_f
\approx
1-n_{\rm ex}
\approx
1-\alpha h^{8/15}e^{-a{\cal A}^2}.
\label{eq:Mf_clean_adiab}
\end{equation}
This prediction can be directly compared with the magnetization from MPS simulations, see the inset in Fig.~\ref{fig:ZNE_M_kink}(d). No additional fitting parameters are introduced, since $\alpha$ and $a$ have already been determined from Fig.~\ref{fig:ZNE_M_kink}(c). Most importantly, the final magnetization depends not only on ${\cal A}$ but also explicitly on $h$, which accounts for the residual dispersion observed in the uniform-bias MPS data in Fig.~\ref{fig:ZNE_M_kink}(d). This dispersion originates from the adiabatic shrinking of the lowest mesonic excitations into single flipped spins by the end of the ramp.

Less deep in the adiabatic regime, higher mesonic states with $l>1$ cannot be neglected. On the one hand, the Landau--Zener formula yields $p_1=p$ in Eq.~\eqref{eq:p}, following from the scaling of the minimal gap near the critical point,
\begin{equation}
\Delta_{\|}\propto h^{8/15}.
\end{equation}
On the other hand, the mesonic gap in Eq.~\eqref{eq:Deltam}, when extrapolated back toward the critical point, scales as $\Delta_{\rm meson}\propto h^{2/3}$. This mean-field exponent differs slightly from the exact value $8/15$ in $\Delta_{\|}$. The $l$-th mesonic mode has energy proportional to $\Delta_{\rm meson}l^{2/3}$, see App.~\ref{app:mesons}. Upon extrapolation back to the critical region by replacing $\Delta_{\rm meson}$ with $\Delta_{\|}$, its excitation probability can be roughly estimated as
\begin{equation}
p_l \approx e^{-a{\mathcal A}^2l^{2/3}}.
\label{eq:pl}
\end{equation}
The normalized probability of observing a domain of size $L=l$ is then
\begin{equation}
P_l =
\frac{p_l}{\sum_{m=1}^{\infty}p_m}.
\label{eq:Pl_normalized}
\end{equation}
The average domain size is therefore
\begin{equation}
\langle L\rangle
=
\sum_{l=1}^{\infty} l P_l
=
\frac{\sum_{l=1}^{\infty} l\,p_l}
{\sum_{l=1}^{\infty}p_l}.
\label{eq:L_average}
\end{equation}
Approximating the sums by integrals gives
\begin{equation}
\langle L\rangle
\propto
\left({\mathcal A}^2\right)^{-3/2}.
\label{eq:L_scaling}
\end{equation}

%%%%%%%%%%%%%%%%%%%%%%%%%%%%%%%%%%%%%%%%%%%%%%%%%%%%%%%%%%%%%%%%%%%%%%%%%%%%%%
\begin{figure}[t]
    \centering
    \includegraphics[width=\linewidth]{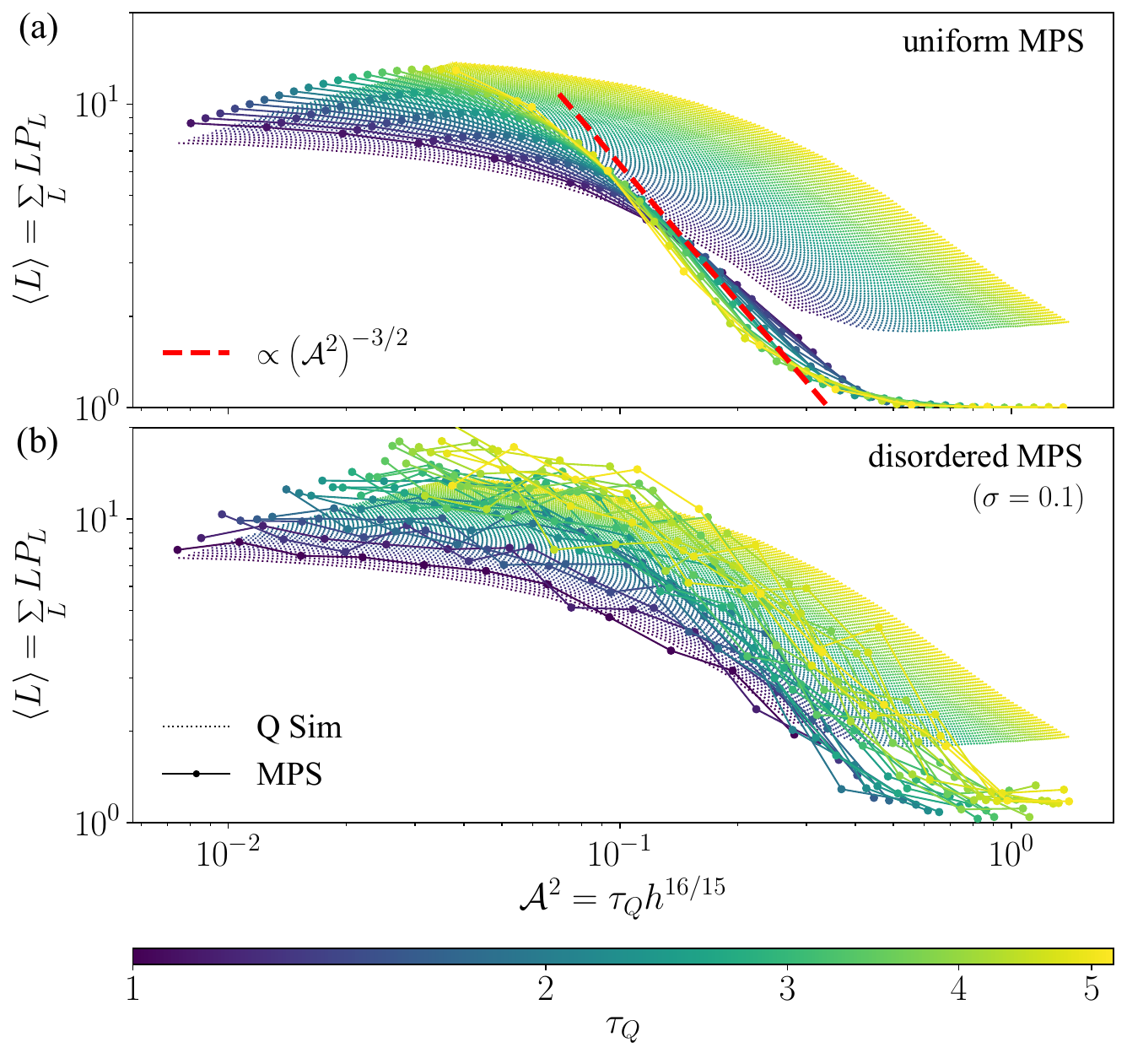}
    \caption{
    \textbf{Domain-size signatures of post-critical meson dynamics and localization.}
    Average minority-domain size $\langle L\rangle=\sum_L L P_L$ as a function of the biased Kibble--Zurek scaling variable ${\cal A}^2=\tau_Q h^{16/15}$, where $P_L$ is the normalized distribution of minority-domain sizes.
    (a) Zero-noise--extrapolated quantum simulation data compared with uniform-bias MPS simulations. The uniform-bias MPS results show an approximate collapse in the post-critical regime and follow the scaling $\langle L\rangle\propto({\cal A}^2)^{-3/2}$ over an intermediate range, consistent with the expected excitation probabilities of higher mesonic bound states.
    (b) The same quantum simulation data compared with disordered MPS simulations with static Gaussian disorder of strength $\sigma=0.1$ in both the coupling $J$ and the longitudinal bias $h$. Disorder broadens the domain-size behavior and disrupts the uniform-bias mesonic scaling, bringing the MPS results closer to the quantum simulation data. The color scale denotes the quench time $\tau_Q$.
    }
    \label{fig:PL}
\end{figure}
%%%%%%%%%%%%%%%%%%%%%%%%%%%%%%%%%%%%%%%%%%%%%%%%%%%%%%%%%%%%%%%%%%%%%%%%%%%%%%

Figure~\ref{fig:PL} compares the resulting domain-size statistics with the quantum simulation data. In the uniform-bias MPS simulations, the average minority-domain size approximately collapses when plotted as a function of $h^{16/15}\tau_Q={\cal A}^2$, consistent with the scaling estimate in Eq.~\eqref{eq:L_scaling}. This behavior reflects post-critical meson dynamics without disorder: the internal mesonic state produced near the critical point determines the final size of the reversed domain.

The zero-noise--extrapolated quantum simulation data do not follow the same collapse as in the case without disorder. Instead, the domain-size statistics show a broader spread, similar to the behavior obtained from disordered MPS simulations with static disorder. This is consistent with the localization picture developed in Sec.~\ref{sec:localization}. Disorder pins the post-critical domain walls and prevents the $l$-th mesonic states from shrinking adiabatically toward the $L=l$ final domains near the end of the ramp.

Taken together, the kink density, magnetization, spatial profiles, and domain-size statistics separate the dynamics into two stages. The number of defects is set mainly by biased Kibble--Zurek creation near the critical point. Their final spatial structure, however, is controlled by post-critical meson dynamics, which in the quantum annealer is interrupted by disorder-induced localization.

%%%%%%%%%%%%%%%%%%%%%%%%%%%%%%%%%%%%%%%%%%%%%%%%%%%%%%%%%%%%%%%%%%%%%%%%%%%%
\section{Conclusion}
\label{sec:conclusion}
%%%%%%%%%%%%%%%%%%%%%%%%%%%%%%%%%%%%%%%%%%%%%%%%%%%%%%%%%%%%%%%%%%%%%%%%%%%%

We have studied nonequilibrium dynamics in a longitudinally biased one-dimensional quantum Ising chain implemented on a 5,564-qubit quantum annealer. The longitudinal bias turns the standard Kibble--Zurek crossing into a problem with two distinct stages. First, defects are created near the rounded quantum critical point, where the relevant competition is between the Kibble--Zurek freeze-out length $\hat\xi$ and the bias-induced magnetic length $\xi_{\|}$. Second, after the system enters the ordered phase, the generated kink--antikink excitations become confined into mesonic bound states and continue to evolve during the remainder of the annealing ramp.

The first main result is that the defect density follows the expected biased Kibble--Zurek/Landau--Zener crossover. After zero-noise extrapolation, the quantum-simulation data collapse as a function of the scaling variable ${\cal A}=\hat\xi/\xi_{\|}$ and agree well with MPS simulations without disorder. This establishes that the near-critical production of defects is controlled by the universal biased Kibble--Zurek mechanism, even in the presence of the finite longitudinal field and the experimental constraints of the annealing platform.

The second main result is that defect counting does not fully characterize the dynamics. The longitudinal magnetization, spatially resolved profiles, and minority-domain statistics reveal post-critical physics that is invisible in the kink density alone. In the uniformly biased Ising chain, the longitudinal field confines the Kibble--Zurek excitations into mesonic bound states. As the transverse field is ramped down, these mesons shrink, and their internal excitation structure determines the final domain-size distribution. This produces residual dependence on the microscopic values of $h$ and $\tau_Q$, beyond the single scaling variable ${\cal A}$, in MPS simulations.

The quantum-annealer data show a different post-critical outcome. While the defect density remains close to the uniformly biased Kibble--Zurek prediction, the magnetization exhibits a stronger collapse as a function of ${\cal A}$, consistent with an earlier freezing of the domain-length imbalance generated near the critical point. Spatial profiles show that kinks preferentially occur at specific positions and are correlated with sharp changes in the local magnetization. Domain-size statistics likewise deviate from the uniform-bias mesonic scaling. These observations indicate that the mesonic evolution is interrupted before the final domain structure is reached.

This interpretation is supported by MPS simulations with static disorder in the longitudinal fields and nearest-neighbor couplings. The disorder leaves the defect density comparatively robust, but localizes the post-critical domain walls, modifies the magnetization, and broadens the domain-size statistics in qualitative agreement with the quantum-simulation data. The comparison therefore identifies static Hamiltonian disorder as a natural mechanism by which the post-critical meson dynamics becomes localized on the annealer.

Our work demonstrates that large-scale coherent quantum annealing can be used to probe more than universal defect creation. In a longitudinal bias, the final spin configurations encode both the creation of critical excitations and their subsequent nonintegrable post-critical dynamics as confined mesons. The combination of zero-noise extrapolated quantum-simulation data, uniform-bias MPS simulations, and disordered MPS simulations allows these two stages to be separated. This establishes a route for using large programmable quantum Ising systems to study the fate of critical excitations, including confinement, mesonic evolution, and localization, in regimes where final-state observables retain information about dynamics beyond the Kibble--Zurek defect density.

%%%%%%%%%%%%%%%%%%%%%%%%%%%%%%%%%%%%%%%%%%%%%%%%%%%%%%%%%%%%%%%%%%%%%%%%%%%%
\section*{Acknowledgments}
\label{sec:acknowledgments}
%%%%%%%%%%%%%%%%%%%%%%%%%%%%%%%%%%%%%%%%%%%%%%%%%%%%%%%%%%%%%%%%%%%%%%%%%%%%
We would like to acknowledge the use of chain embeddings provided by Dennis Willsch in our quantum simulations.
F.B. acknowledges support from the National Science Centre (NCN), Poland, under project 2020/38/E/ST3/00150. 
M.R. and J.D. acknowledge support from the National Science Centre (NCN), Poland, under project 2024/55/B/ST3/00626. 
M.R and J.D. were also supported by a grant from the Priority Research Area DigiWorld under the Strategic Programme Excellence Initiative at Jagiellonian University.
J.V. acknowledges support from the project Jülich UNified Infrastructure for Quantum computing (JUNIQ) that has received funding from the German Federal Ministry of Education and Research (BMBF) and the Ministry of Culture and Science of the State of North Rhine-Westphalia, financial support from ARIS, P1-0040 Nonequilibrium Quantum System Dynamics, P1-0416 Physics of Quantum Technologies, J1-70063 Quantum Simulation of Non-equilibrium Phenomena on Quantum Devices, and ERC AdG HIMMS – Hidden metastable mesoscopic states in quantum materials. The authors gratefully acknowledge the Jülich Supercomputing Centre (https://www.fzjuelich.de/ias/jsc) for funding this project by providing computing time on the D-Wave Advantage™ System JUPSI through the Jülich UNified Infrastructure for Quantum computing (JUNIQ).

%%%%%%%%%%%%%%%%%%%%%%%%%%%%%%%%%%%%%%%%%%%%%%%%%%%%%%%%%%%%%%%%%%%%%%%%%%%%
\section*{Data availability}
\label{sec:data}
%%%%%%%%%%%%%%%%%%%%%%%%%%%%%%%%%%%%%%%%%%%%%%%%%%%%%%%%%%%%%%%%%%%%%%%%%%%%
Quantum simulation data obtained from the quantum annealer and matrix product state simulation data, along with the corresponding figure plotting scripts, were deposited to Zenodo and are available at the following URL: \url{https://doi.org/10.5281/zenodo.21339733}.

%%%%%%%%%%%%%%%%%%%%%%%%%%%%%%%%%%%%%%%%%%%%%%%%%%%%%%%%%%%%%%%%%%%%%%%%%%%%

%%%%%%%%%%%%%%%%%%%%%%%%%%%%%%%%%%%%%%%%%%%%%%%%%%%%%%%%%%%%%%%%%%%%%%%%%%%%
\bibliographystyle{apsrev4-2}
\bibliography{KZref.bib}

%%%%%%%%%%%%%%%%%%%%%%%%%%%%%%%%%%%%%%%%%%%%%%%%%%%%%%%%%%%%%%%%%%%%%%%%%%%%
\appendix
%%%%%%%%%%%%%%%%%%%%%%%%%%%%%%%%%%%%%%%%%%%%%%%%%%%%%%%%%%%%%%%%%%%%%%%%%%%%

%%%%%%%%%%%%%%%%%%%%%%%%%%%%%%%%%%%%%%%%%%%%%%%%%%%%%%%%%%%%%%%%%%%%%%%%%%%%
\section{Quantum annealing parameter sweep}
\label{app:parameter_sweep}
%%%%%%%%%%%%%%%%%%%%%%%%%%%%%%%%%%%%%%%%%%%%%%%%%%%%%%%%%%%%%%%%%%%%%%%%%%%%

Here we summarize the parameter sweep used for the quantum annealing experiments. For each value of the longitudinal bias $h$ in the range $[0.01 , 0.015, 0.023, 0.035, 0.053, 0.081, 0.123, 0.187, 0.285,\\
       0.433, 0.658, 1.]$, the experiment was repeated over a two-dimensional grid of hardware energy scaling factors $E$ and annealing times $t_a$. The scaling factor $E$ rescales the programmed longitudinal fields and Ising couplings according to $h\rightarrow Eh$ and $J\rightarrow EJ$, or equivalently $B(s)\rightarrow E B(s)$ in the effective annealing schedule. This sweep provides the set of finite-noise data points used for the zero-noise extrapolation described in Sec.~\ref{sec:annealer_realization}.

\begin{figure}[t]
    \centering
    \includegraphics[width=\linewidth]{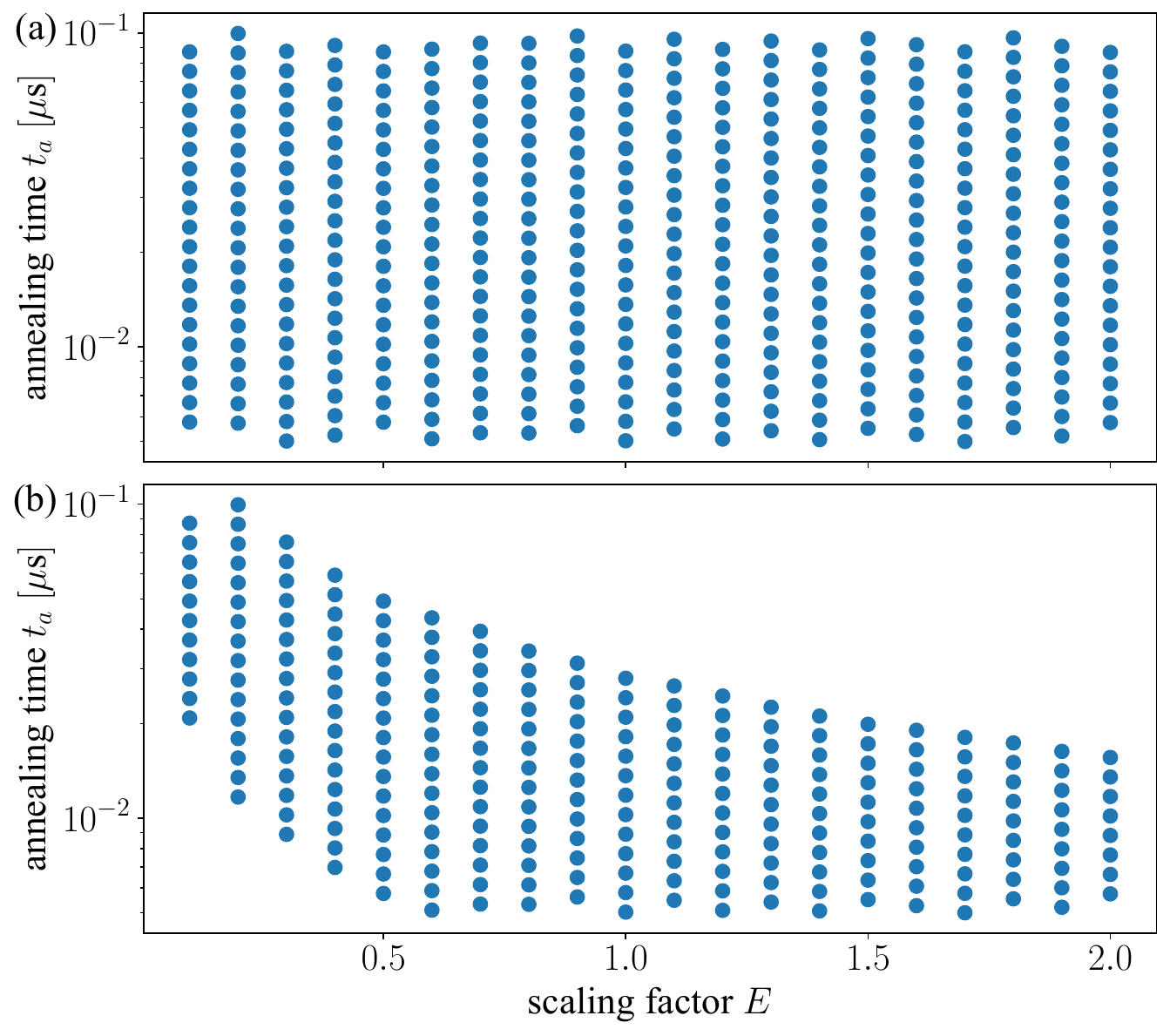}
    \caption{
    \textbf{Annealing-time sweep used in the quantum simulation.}
    (a) Hardware annealing times $t_a$ used for different energy scaling factors $E$.
    (b) The same sweep after conversion to the dimensionless quench time $\tau_Q=t_a/t_Q(E)$ and restriction to the window $5\times10^{-1}\leq\tau_Q\leq5\times10^0$ used in the zero-noise extrapolation.
    Because $t_Q(E)$ depends on the energy scaling factor, a rectangular sweep in hardware time $t_a$ maps to an $E$-dependent range of dimensionless quench times.
    }
    \label{fig:anneal_time_sweep}
\end{figure}

The raw hardware sweep was performed over a broad range of annealing times for each value of $E$, as shown in Fig.~\ref{fig:anneal_time_sweep}(a). Since the relevant dynamical variable is the dimensionless quench time $\tau_Q=t_a/t_Q(E)$, the same hardware annealing-time grid corresponds to different values of $\tau_Q$ at different energy scales. After converting from $t_a$ to $\tau_Q$, we retained the data in the range
\begin{equation}
5\times10^{-1}\leq \tau_Q \leq 5\times10^0,
\end{equation}
as shown in Fig.~\ref{fig:anneal_time_sweep}(b). This filtering gives a common dynamical window across the energy-scale sweep and ensures that the zero-noise extrapolation compares data at fixed $\tau_Q$ while varying the noise parameter $\lambda$ through $E$.

For each target value of $\tau_Q$, the measured observables were grouped by the corresponding energy scaling factors $E$ and converted to the dimensionless noise parameter $\lambda=t_Q(E)/t_Q(1)$. The zero-noise estimates of the magnetization, defect density, spatial profiles, and domain-size observables were then obtained by extrapolating the measured data to $\lambda=0$.

%%%%%%%%%%%%%%%%%%%%%%%%%%%%%%%%%%%%%%%%%%%%%%%%%%%%%%%%%%%%%%%%%%%%%%%%%%%%%%
%%%%%%%%%%%%%%%%%%%%%%%%%%%%%%%%%%%%%%%%%%%%%%%%%%%%%%%%%%%%%%%%%%%%%%%%%%%%%%

%%%%%%%%%%%%%%%%%%%%%%%%%%%%%%%%%%%%%%%%%%%%%%%%%%%%%%%%%%%%%%%%%%%%%%%%%%%%%%
\begin{figure}[t]
    \centering
    \includegraphics[width=\linewidth]{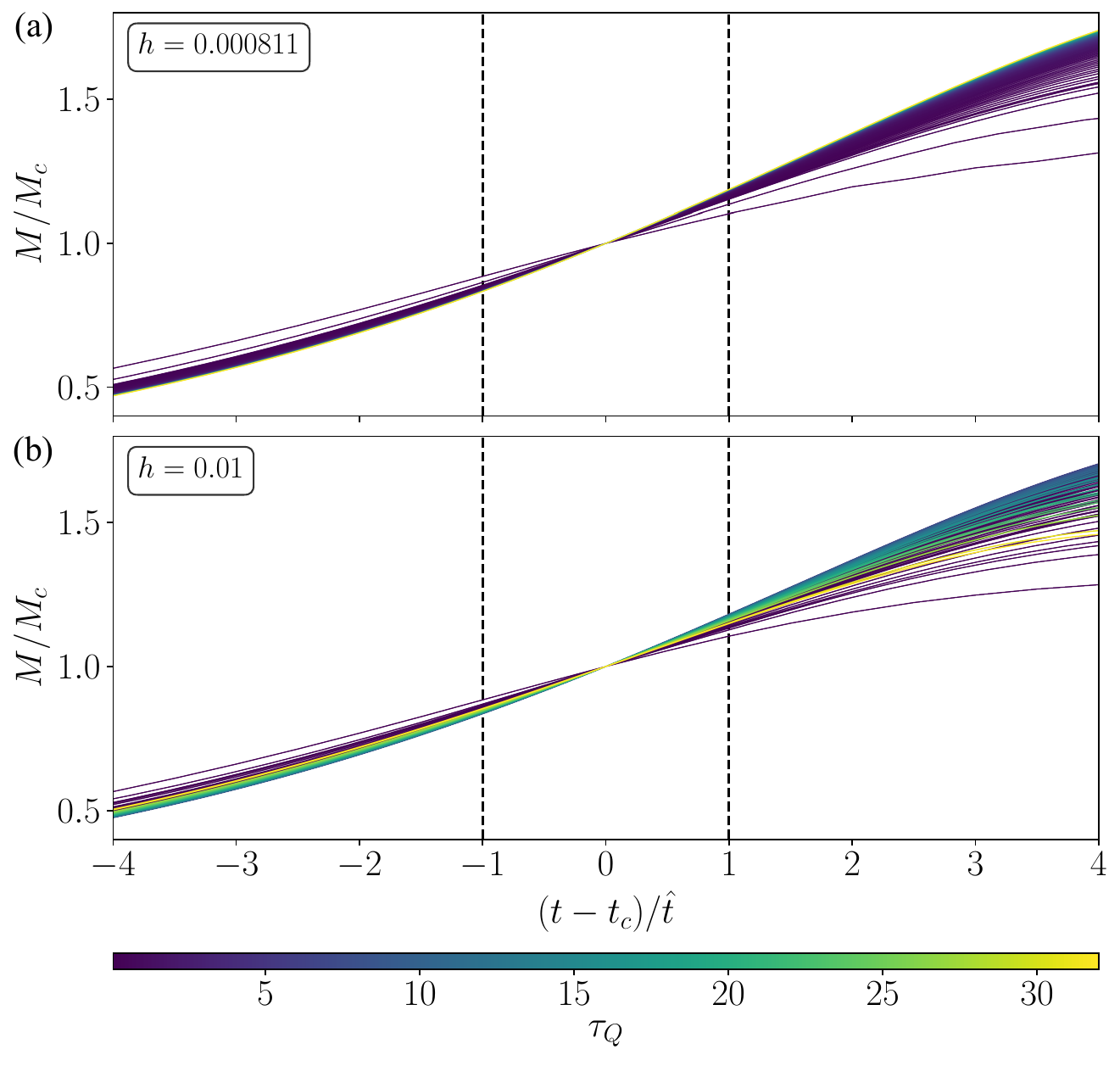}
    \caption{
    \textbf{Critical growth of the bias-induced magnetization.}
    Uniform-bias MPS simulations of the rescaled magnetization $M/M_c$ as a function of the Kibble--Zurek time variable $(t-t_c)/\hat t$ for two weak longitudinal biases, (a) $h=0.000811$ and (b) $h=0.01$. The color scale denotes the quench time $\tau_Q$.
    For both values of $h$, the curves approach a common scaling form as $\tau_Q$ increases, demonstrating that the magnetization dynamics in the interval shown is governed by the Kibble--Zurek critical regime.
    The vertical dashed lines mark $t=t_c\pm\hat t$, with $\hat t=\sqrt{\tau_Q}$.
    Rather than remaining strictly frozen through the impulse regime, the magnetization continues to grow due to critical ordering, increasing by a factor $\kappa\gtrsim3.3$ between the initial freeze-out point and the end of the critical window. This amplification is used in the main text to relate the bias-induced magnetization to the final imbalance between domains aligned and anti-aligned with the longitudinal field.
    }
    \label{fig:MoverMc}
\end{figure}
%%%%%%%%%%%%%%%%%%%%%%%%%%%%%%%%%%%%%%%%%%%%%%%%%%%%%%%%%%%%%%%%%%%%%%%%%%%%%%

%%%%%%%%%%%%%%%%%%%%%%%%%%%%%%%%%%%%%%%%%%%%%%%%%%%%%%%%%%%%%%%%%%%%%%%%%%%%
\section{Critical coarsening of the magnetization}
\label{app:sonic_M}
%%%%%%%%%%%%%%%%%%%%%%%%%%%%%%%%%%%%%%%%%%%%%%%%%%%%%%%%%%%%%%%%%%%%%%%%%%%%

Figure \ref{fig:MoverMc} shows the rescaled magnetization $M/M_c$ as a function of the rescaled time $(t-t_c)/\hat t$. The rescaling is performed according to the KZ theory. For small values of the bias $h$, the curves collapse onto a common scaling function, confirming the theoretical predictions for the nonadiabatic stage of the time evolution. One can also see that, during this stage, the magnetization does not freeze out completely but continues to increase due to critical coarsening. Consequently, the ratio $M/M_c$ can grow by a factor of several, quantified by $\kappa$. This increase in magnetization reflects the growth of ferromagnetic domains aligned with the bias at the expense of anti-aligned domains.

%%%%%%%%%%%%%%%%%%%%%%%%%%%%%%%%%%%%%%%%%%%%%%%%%%%%%%%%%%%%%%%%%%%%%%%%%%%%
\section{Evolution of mesonic states}
\label{app:mesons}
%%%%%%%%%%%%%%%%%%%%%%%%%%%%%%%%%%%%%%%%%%%%%%%%%%%%%%%%%%%%%%%%%%%%%%%%%%%%

Here we extend the theory of Ref.~\onlinecite{Lagnese_Wilczek} to the post-critical dynamics of mesonic states. The Hamiltonian \ref{eq:HDW} is
\be
H = 
- \Gamma \sum_i \sigma^x_i 
- {\mathcal J}\left( h \sum_i \sigma^z_i + 
                     \sum_i \sigma^z_i \sigma^z_{i+1} \right)  .
\ee
In the absence of the bias, $h=0$, the dispersion relation for the fermionic Bogoliubov quasiparticles is
\bea
\omega(k)=2\sqrt{(\Gamma-{\mathcal J}\cos k)^2+{\mathcal J}^2\sin^2k}\approx \omega_0 + \frac{k^2}{2m}.
\eea
Here the quasiparticle gap is $\omega_0=2({\mathcal J}-\Gamma)$ and the quasiparticle mass is $m=\frac{({\mathcal J}-\Gamma)}{2{\mathcal J}\Gamma}$. The long-wavelength (continuum) approximation requires
$k^2 \ll \frac{2({\mathcal J}-\Gamma)^2}{{\mathcal J}\Gamma}$.
The Hamiltonian for a pair of fermionic quasiparticles is
\be
H=\omega(p_1)+\omega(p_2)+2{\mathcal J}h M_s|x_1-x_2|.
\ee
Here $M_s=[1-(\Gamma/{\mathcal J})^2]^{1/8}$ denotes the spontaneous magnetization. The linear confining potential is accurate when the separation between the quasiparticles is much larger than their width, which is given by the healing length $\frac{\sqrt{{\mathcal J}\Gamma}}{({\mathcal J}-\Gamma)}$. This is the same condition required for the continuum approximation.

Introducing the center-of-mass and relative coordinates,
$X=(x_1+x_2)/2$, $P=p_1+p_2$, $x=x_1-x_2$, and $p=(p_1-p_2)/2$, we obtain
\be
H=2\omega_0+\frac{P^2}{4m}+\frac{p^2}{m}+2{\mathcal J}hM_s|x|.
\ee
For a total eigenenergy $E^{\rm tot}_l=2\omega_0+\frac{P^2}{4m}+E_l$, the relative wavefunction satisfies the Schr\"odinger equation
\be
E_l \psi_l(x) = -\frac{1}{m} \partial_x^2 \psi_l(x) + 2{\mathcal J}hM_s x \psi_l(x).
\ee
Since the fermionic wavefunction $\psi_l(x)$ must be antisymmetric, it is sufficient to consider $x>0$ with the boundary condition $\psi_l(0)=0$. The eigenmodes are expressed in terms of the Airy function:
\bea
&&
\psi_l(x,\xi) = N_l~\xi^{-1/2} {\rm Ai}\left(x/\xi-z_l\right),
\label{eq:airy} \\
&&
E_l=\frac{|z_l|}{\xi^2 m}.
\label{eq:El}
\eea
Here $\xi=(2{\mathcal J}hmM_s)^{-1/3}$ denotes the meson size, $z_l$ is the $l$-th zero of the Airy function, and $N_l=|{\rm Ai}'(z_l)|^{-1}$ is the normalization constant. 

%%%%%%%%%%%%%%%%%%%%%%%%%%%%%%%%%%%%%%%%%%%%%%%%%%%%%%%%%%%%%%%%%%%%%%%%%%%%%%
\begin{figure}[t]
    \centering
    \includegraphics[width=\linewidth]{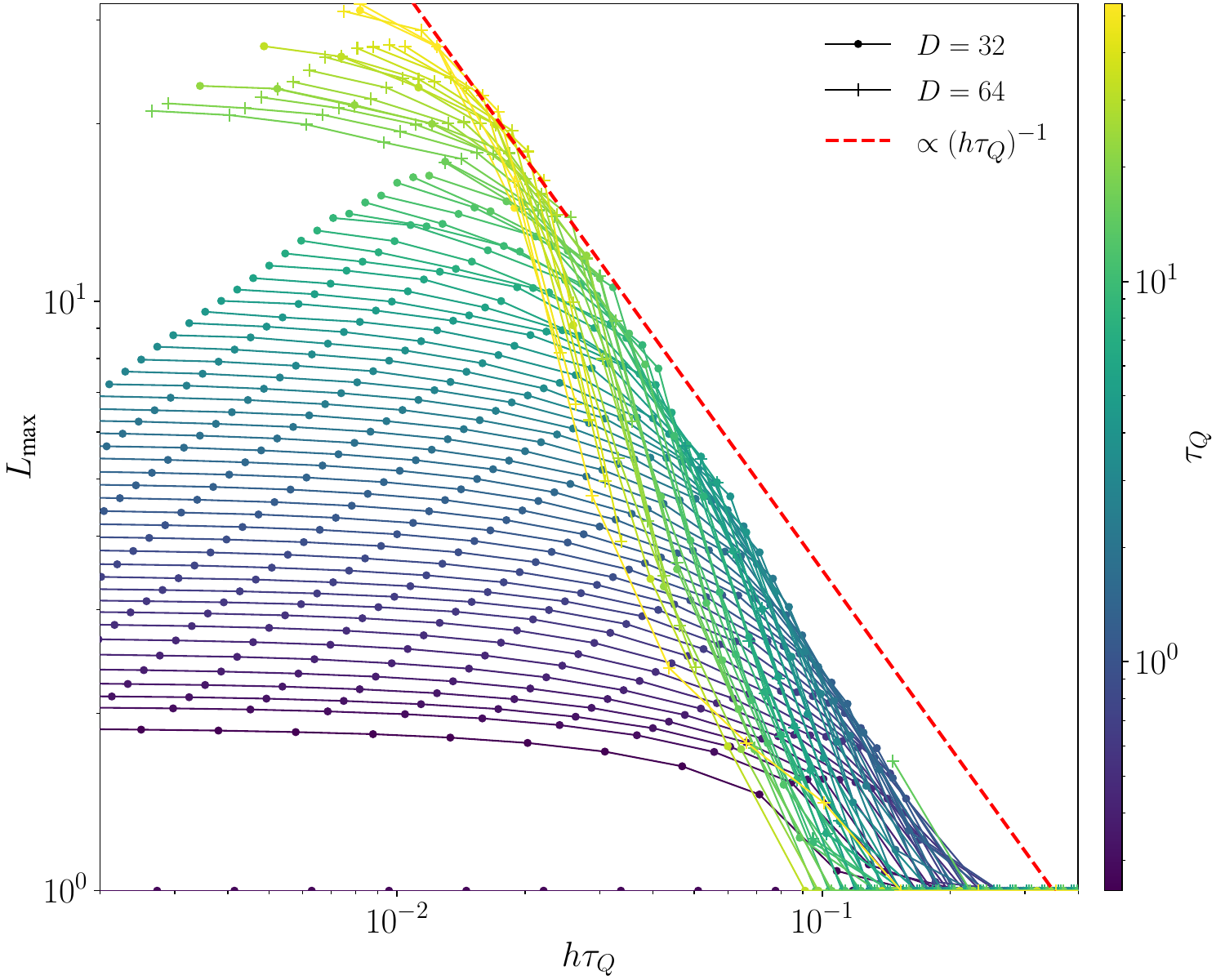}
    \caption{
    \textbf{Freeze-out of the meson size during the post-critical ramp.}
    Most likely minority-domain size $L_{\rm max}$ obtained from uniform-bias MPS simulations as a function of $h\tau_Q$. Results are shown for bond dimensions $D=32$ and $D=64$, with the color scale denoting the quench time $\tau_Q$.
    The dashed line indicates the scaling $L_{\rm max}\propto(h\tau_Q)^{-1}$ predicted from the freeze-out of the shrinking meson wavefunction during the late part of the ramp.
    This scaling is observed in an intermediate regime between the Kibble--Zurek regime at small ${\cal A}=\sqrt{\tau_Q}h^{8/15}$, where the domain size is set primarily by critical defect creation, and the adiabatic regime at large $\tau_Q h^{16/15}$, where the mesonic states evolve into short lattice-scale domains.
    }
    \label{fig:peak_scaled}
\end{figure}
%%%%%%%%%%%%%%%%%%%%%%%%%%%%%%%%%%%%%%%%%%%%%%%%%%%%%%%%%%%%%%%%%%%%%%%%%%%%%%

We note that for $\Gamma \ll {\mathcal J}$ the size decreases with decreasing $\Gamma$ as
\be
\xi = \left(\frac{\Gamma}{{\mathcal J}h}\right)^{1/3},
\label{eq}
\ee
while the eigenenergies shrink as
\be
E_l=2|z_l|\Gamma^{1/3}({\mathcal J}h)^{2/3}.
\ee
Near the end of the ramp, when $\Gamma \ll {\mathcal J}h$, the predicted size becomes smaller than the lattice spacing $1$, and the continuum approximation breaks down. Upon entering this regime, the $l$-th mesonic state evolves adiabatically into a domain of $l$ reversed spins. The domain eigenstates are separated by energy gaps that saturate at ${\mathcal J}h$.

For a time-dependent ramp, transitions between eigenmodes $l_1$ and $l_2$ are governed within adiabatic perturbation theory by the rates
\be
\bra{l_1}\frac{d}{dt}\ket{l_2} = -\gamma N_{l_1l_2}.
\ee
Here the transition rate is
\be
\gamma=\frac{\dot \xi}{\xi} = \frac13 \left( \frac{\dot\Gamma}{\Gamma} - \frac{\dot {\mathcal J}}{{\mathcal J}} \right),
\ee
and the matrix element is
\be
N_{l_1l_2} = N_{l_1} N_{l_2} \int_0^\infty ydy~{\rm Ai}(y-y_{l_1}) {\rm Ai}'(y-y_{l_2}).
\ee
For $l_1=1$ and $l_2=2,3,4,\ldots$, the corresponding values are $1.1198$, $-0.186$, $0.068$, $\ldots$, respectively. The transition rate can be compared with the corresponding transition frequency
\be
\Delta_{l_1l_2} = 
E_{l_1}-E_{l_2} \approx 
2|z_{l_1}-z_{l_2}|\Gamma^{1/3}({\mathcal J}h)^{2/3}.
\label{eq:E-E}
\ee
The adiabaticity condition is $|\gamma N_{l_1l_2}|\ll|\Delta_{l_1l_2}|$.
Near the end of the D-Wave ramp, $\Gamma(t)$ decays exponentially as $\Gamma(t)\approx\Gamma_0e^{-\alpha t/t_a}$, while ${\mathcal J}(t)$ approaches its final value linearly as ${\mathcal J}(t)\approx {\mathcal J}_f-\beta {\mathcal J}_f\frac{t_a-t}{t_a}$. The corresponding rate is $|\gamma|=\frac13(\alpha-\beta)/t_a$. For slow ramps, the adiabaticity condition breaks down when $\Gamma(t)$ falls below
\be
\tilde\Gamma \propto ({\mathcal J}_fh)^{-2} t_a^{-3}.
\label{eq:tildeGamma}
\ee
At $\tilde\Gamma$, the meson size freezes at
\be
\tilde \xi \propto \frac{1}{ {\mathcal J}_f h t_a } = \frac{1}{ {\mathcal J}_f t_Q(1)} (h \tau_Q)^{-1}.
\label{eq:tildexi}
\ee
This expression remains valid provided that it predicts $\tilde\xi\gg1$. In the opposite case the $l$-th mesonic mode evolves adiabatically into a domain of $L=l$ reversed spins.

In the adiabatic regime the lowest mesonic states, $l=1$, are most likely to get excited near the critical point with the probability \eqref{eq:p}. As long as $\tilde\xi\gg1$, the probability distribution for the domain size $L$ reads 
\be 
P_L = \psi_1^2\left( L,\tilde\xi\right) \propto {\rm Ai}^2\left(L/\tilde\xi\right).
\label{eq:PLapp}
\ee 
The distribution is accurate near its peak at
\be 
L_{\rm max} \propto \tilde\xi.
\label{eq:Lmax}
\ee
This peak position is confirmed by Fig. \ref{fig:peak_scaled}. In a narrow intermediate regime between the Kibble–Zurek (KZ) regime at small $\mathcal{A}=\tau_Q h^{16/15}$ and the adiabatic regime at large $\tau_Q h^{16/15}$, it follows the power law $L_{\rm max} \propto (h\tau_Q)^{-1}$.

\end{document}